\documentclass{article}
\usepackage{spconf,amsmath,graphicx,subfigure, amsfonts}
\usepackage{enumerate}
\usepackage[english]{babel}
\usepackage{indentfirst}

\title{DCTNet and PCANet for acoustic signal feature extraction}
\name{Yin Xian$^{\star}$, Andrew Thompson$^{\dagger}$, Xiaobai Sun$^{\ddag}$, Douglas Nowacek$^{\star}$ and Loren Nolte$^{\star}$}
\address{$^{\star}$ Department of Electrical and Computer Engineering, Duke University, Durham, NC 27708, USA \\
$^{\dagger}$ Mathematical Institute, University of Oxford, Oxford OX2 6GG, UK\\
$^{\ddag}$ Department of Computer Science, Duke University, Durham, NC 27708, USA \\
\{yin.xian,~xiaobai.sun,~doug.nowacek,~loren.nolte@duke.edu\}, \{thompson@maths.ox.ac.uk\}\
} 

\begin{document}
%
\maketitle
\begin{abstract}
We introduce the use of DCTNet, an efficient approximation and alternative to PCANet, for acoustic signal classification. In PCANet, the eigenfunctions of the local sample covariance matrix (PCA) are used as filterbanks for convolution and feature extraction. When the eigenfunctions are well approximated by the Discrete Cosine Transform (DCT) functions, each layer of of PCANet and DCTNet is essentially a time-frequency representation. We relate DCTNet to spectral feature representation methods, such as the the short time Fourier transform (STFT), spectrogram and linear frequency spectral coefficients (LFSC). Experimental results on whale vocalization data show that DCTNet improves classification rate, demonstrating DCTNet's applicability to signal processing problems such as underwater acoustics.
\end{abstract}
\begin{keywords}
Convolutional network, PCA, DCT, filterbanks, acoustic perception, spectral clustering, whale vocalizations.
\end{keywords}
\section{Introduction}
\label{sec:intro}
The power of multi-layer convolutional networks for learning features has been established in audio signal processing, such as speech recognition and music classification~\cite{lecun1995convolutional, hinton2012deep, lee2009unsupervised}. The idea of a convolutional network is to convolve signals with filters, and use the obtained features for classification. The scattering transform, proposed by Mallat~\cite{mallat2012group}, employed pre-specified wavelets as filters, and obtained state of the art result on some music and speech datasets~\cite{anden2011multiscale,anden2014deep}. The structure of the scattering transform is similar to a cascade of constant-Q or mel-filter banks~\cite{anden2011multiscale}, and the scattering transform can capture useful spectral contents of acoustic signals.

The spectral content of acoustic signals is often analyzed using
time-frequency energy distributions with a particular frequency scale. For example, the popular mel-frequency cepstral coefficients (MFCC) are based on the mel-scale, a human perceptual frequency scale~\cite{stevens1937scale}. The MFCC is a DCT of the log energy of the mel-frequency spectral coefficients (MFSC)~\cite{rabiner1993fundamentals}, which were shown to be closely related to scattering transform coefficients~\cite{anden2011multiscale,anden2014deep}.

PCANet~\cite{chan2014pcanet} was recently proposed for image classification. In this framework, filters are learned from the data as principal components at the local ``image patch'' level. PCANet was shown to match and in some cases improve upon state of the art performance in a variety of image classification benchmarks.

\begin{figure}[htb]
\centering
\includegraphics[height=50mm,width=90mm]{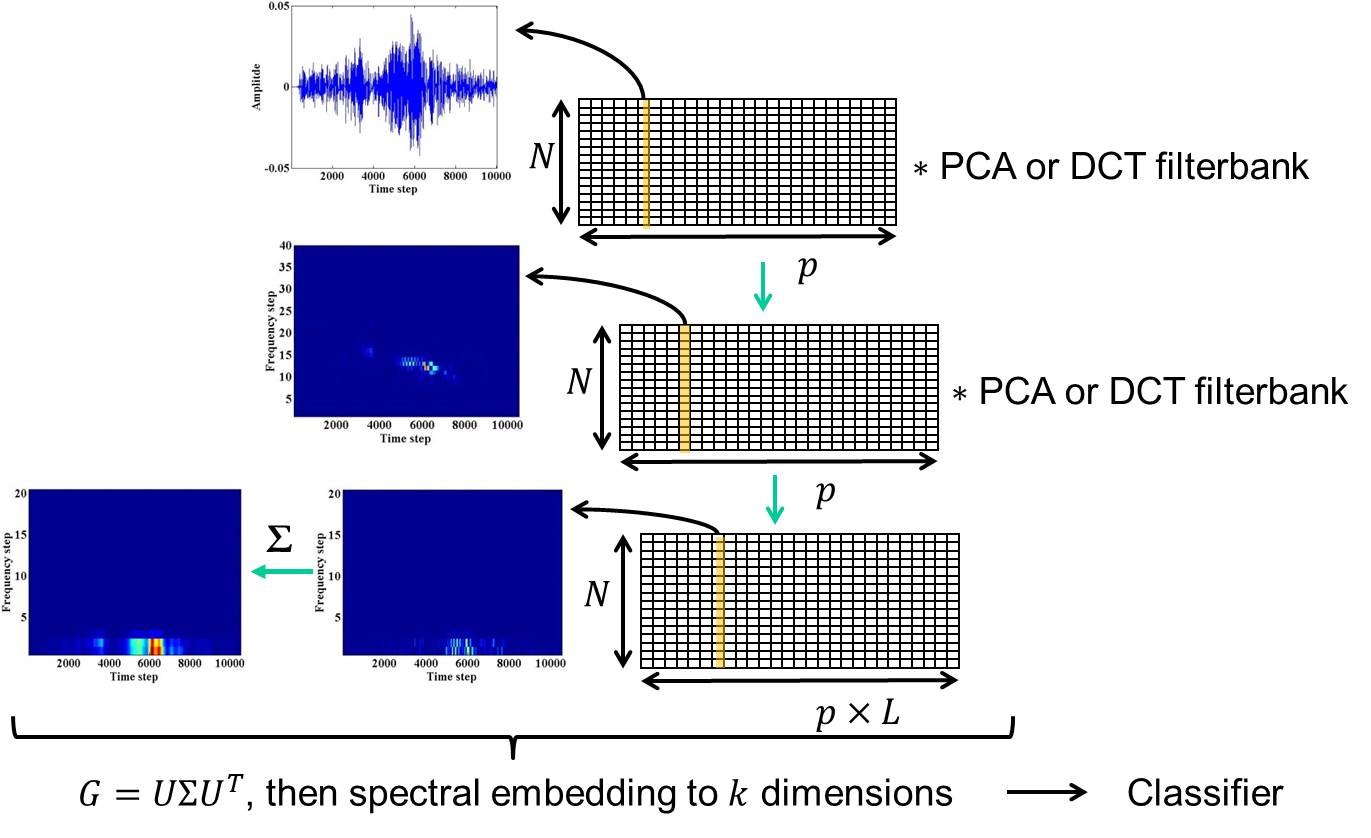}
\vspace{-10pt}
\caption{\small{PCANet and DCTNet Process. The input is the time series of an acoustic signal. After convolving with DCT or PCA filterbanks, we have the short time DCT or short time PCA of the signal. Summing and averaging the second layer's outputs, we have linear scale spectral coefficients, and we use them for spectral clustering and classification.}}
\label{fig:process}
\end{figure}

In this paper, we translate the PCANet framework into the world of acoustic signal processing and relate it to spectral feature representations. The PCANet filters are obtained as the eigenvectors of a local covariance matrix, which is a Toeplitz matrix, and so the resulting filters can be approximated by the Discrete Cosine Transform (DCT) basis functions~\cite{ahmed1974discrete,sanchez1995diagonalizing}. We thus introduce the use of DCTNet for acoustic signal classification, in which PCA filters are simply replaced with fixed DCT filters. We relate DCTNet to spectral feature representation methods for acoustic signals, such as the short time Fourier transform (STFT), spectrogram and linear frequency spectral coefficients (LFSC). In particular, each DCTNet layer is essentially a short time DCT. The process of our PCANet and DCTNet is shown in Fig.~\ref{fig:process}. More technical details can be found in Section~\ref{sec:signal} to Section~\ref{sec:lfsc} and also in the dissertation of Xian~\cite{xian2015whale}.

DCTNet is of interest as an alternative to the scattering transform for the following reasons. Firstly, DCTNet features explicitly give frequency information, which is likely beneficial for many acoustic signals, whereas scattering transform coefficients capture scale information. Secondly, the DCT is a popular tool in audio signal coding~\cite{princen1986analysis}, for example in MP3, and so DCTNet provides a natural way to incorporate software and hardware in these contexts into a deep learning framework.

We note that Ng and Teoh~\cite{ng2015dctnet} recently introduced the use of a DCTNet variant for image feature extraction, applying block-wise 2D convolution, histogramming and binary hashing. In contrast, we adopt DCTNet for acoustic signals, using an entirely different post-processing strategy, and we also provide insight into the different time-frequency content revealed by each layer of DCTNet.

We present experiments using the DCLDE whale vocalization data~\cite{dclde}. Results show that DCTNet improved results in classification tasks, suggesting that DCTNet is an attractive tool for acoustic signal processing problem, such as underwater acoustics.

\section{The DCT Approximation for PCANet Eigenfunctions}
\label{sec:signal}
In the PCANet framework~\cite{chan2014pcanet}, filterbanks are obtained as
the eigenfunctions of the local covariance matrix. Given a signal $\bold{x}=\small{(x(1), x(2),\cdots, x(N))}$, we construct the Hankel matrix
\small
\begin{align*}
\textbf{X}=\left[\begin{array}{cccc}
x(1) & x(2) &\cdots & x(N-M+1) \\
x(2) & x(3) &\cdots & x(N-M+2) \\
\vdots & \vdots & \ddots & \vdots \\
x(M) & x(M+1) & \cdots & x(N)
\end{array}\right],
\end{align*}
\normalsize
where $M<N$. When the first column $\small{x(1), \cdots, x(M)}$ and the last column $x(N-M+1),\cdots, x(N)$ are zeros, letting $\rho_j=\sum\limits_{i} x(i)x(i+j)$, the sample covariance matrix
\small
\begin{align*}
\bold{XX^{T}}=\left[\begin{array}{ccccc}
1 & \rho_1 & \rho_2 &\cdots & \rho_{M-1} \\
\rho_1 & 1 & \rho_1 &\cdots & \rho_{M-2} \\
\vdots & \vdots & \vdots & \ddots & \vdots \\
\rho_{M-1} & \rho_{M-2} & \rho_{M-3} &\cdots & 1
\end{array}\right]
\end{align*}
\normalsize
is a Toeplitz matrix.

\begin{figure}[!ht]
\centering
\subfigure[PCA eigenfunctions 1-4]{\includegraphics[width=.45\linewidth,height=2.3cm]{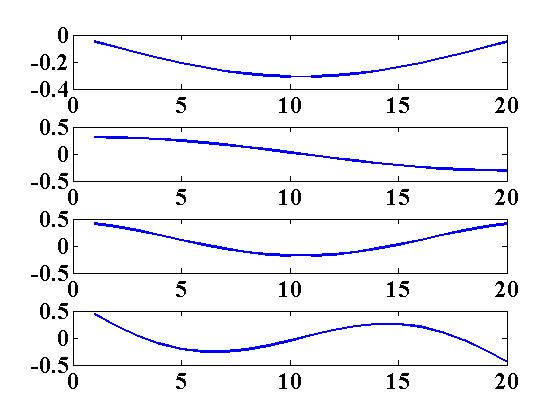}}
\subfigure[PCA eigenfunctions 5-8]{\includegraphics[width=.45\linewidth,height=2.3cm]{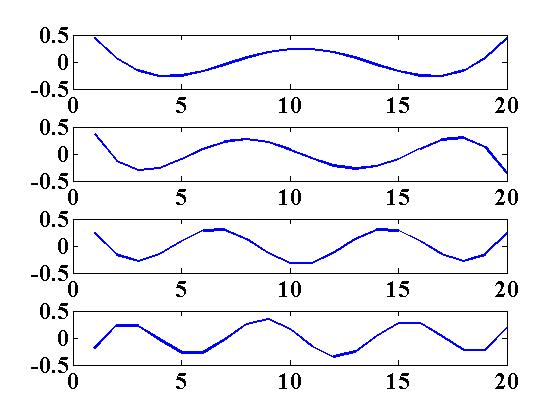}}
\\
\vspace{-6pt}
\centering
\subfigure[DCT eigenfunctions 1-4]{\includegraphics[width=.45\linewidth,height=2.3cm]{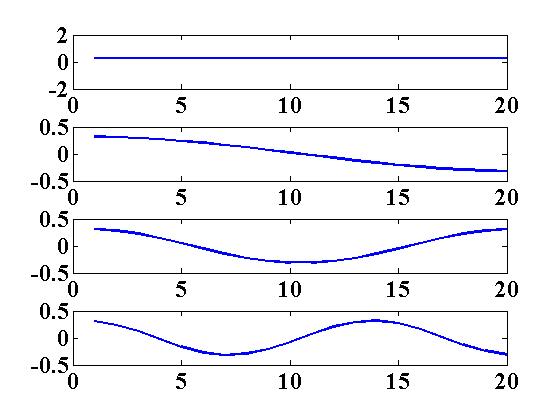}} 
\subfigure[DCT eigenfunctions 5-8]{\includegraphics[width=.45\linewidth,height=2.3cm]{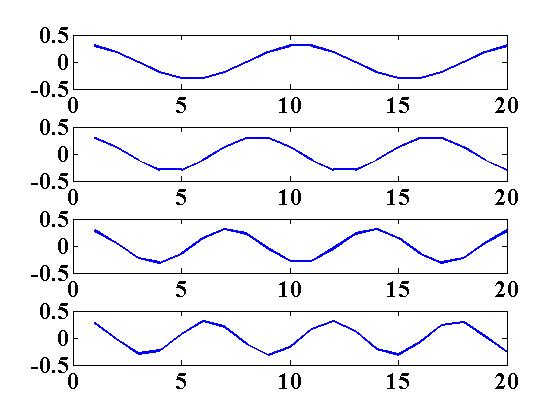}}
\vspace{-6pt}
\caption{Comparison of the top eight DCT eigenfunctions and PCA eigenfunctions for a single whale vocalization}
\label{fig:eigenfunctions_comp}
\end{figure}

When the autocorrelation of the signal decays fast, the discrete cosine transform (DCT) basis functions can well approximate the eigenfunctions of the Toeplitz matrix~\cite{ahmed1974discrete,sanchez1995diagonalizing}. A comparison of the top eight eigenfunctions of DCT and PCA for a single whale vocalization are shown in Fig.~\ref{fig:eigenfunctions_comp}.

\begin{figure}[!ht]
\centering
\subfigure[Autocorrelation of signal]{\includegraphics[width=.45\linewidth,height=2.55cm]{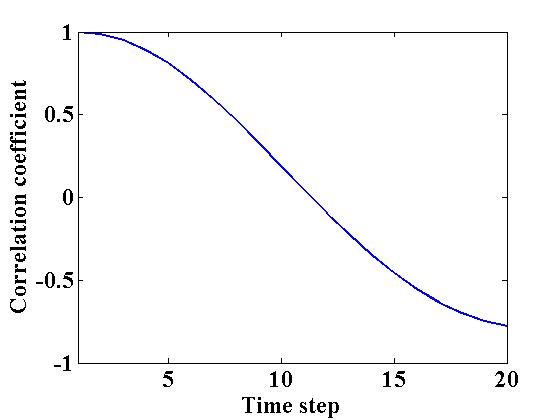}}
\subfigure[Eigenfunctions correlation]{\includegraphics[width=.45\linewidth,height=2.55cm]{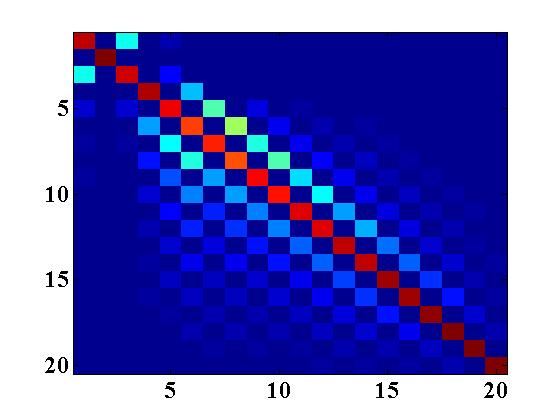}}
\vspace{-6pt}
\caption{Signal autocorrelation and eigenfunctions correlation}
\label{fig:eigenfunctions_comp2}
\end{figure}

The autocorrelation of the signal and the correlation between the DCT and PCA eigenfunctions are shown in Fig.~\ref{fig:eigenfunctions_comp2}. We can see that the approximation of DCT to PCA eigenfunctions depends upon the structure of the data. An error bound for using DCT eigenfunctions to diagonalize the Toeplitz matrix has been given in terms of the autocorrelation coefficients of the signal~\cite{sanchez1995diagonalizing}.

We can also interpret the DCT approximation in a Toeplitz-Fourier framework, since a Toeplitz matrix $T$ can be represented as the sum of a circulant matrix $C$ and a skew-circulant matrix $S$, that is $T=C+S$. The eigenfunctions of a circulant matrix are well known to be Fourier eigenfunctions. We can view the skew-circulant matrix as an error, and the energy of the circulant matrix within the given Toeplitz matrix can be optimized by applying Chan's optimal circulant preconditioner~\cite{chan1988optimal}, that is $C=\arg\min\limits_{C'}||C'-T||_{F}^2$.  By this argument, the eigenfunctions are well approximated by Fourier basis functions, and consequently by the closely related DCT basis functions.

\section{Short time PCA and short time DCT}
\label{sec:stft}
The discrete time STFT can be written as~\cite{oppenheim1978applications}:
\small
\begin{align*}
X(m,\omega)&=\sum\limits_{n=-\infty}^{\infty}x(m-n)w(n)\exp(-j\omega (m-n)) \\
&=\exp(-j\omega m)\sum\limits_{n=-\infty}^{\infty}(w(n)\exp(j\omega n))x(m-n) \\
&=\exp(-j\omega m)[(w(m)\exp(j\omega m))*x(m)]
\end{align*}
\normalsize
where $w$ is a window function and $\omega$ is the angular frequency. It can be viewed as the modulation of a band-pass filter. For PCANet and DCTNet, we replace the Fourier basis functions $\{\exp(j\omega m)\}$ with PCA eigenfunctions and DCT eigenfunctions, obtaining a short time PCA and short time DCT of the signal respectively.

\begin{figure}[!ht]
\centering
\subfigure[DCTNet with window size 256]{\includegraphics[width=.45\linewidth,height=2.8cm]{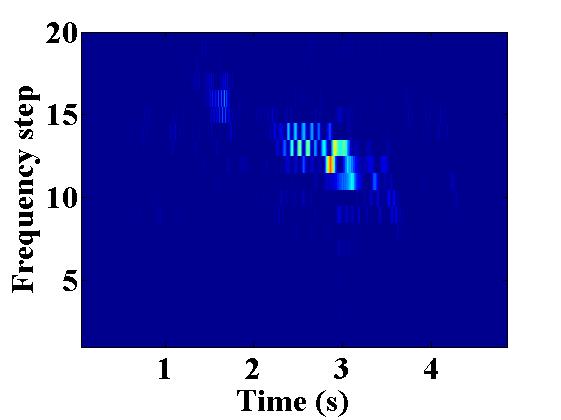}}
\subfigure[DCTNet with window size 20]{\includegraphics[width=.45\linewidth,height=2.8cm]{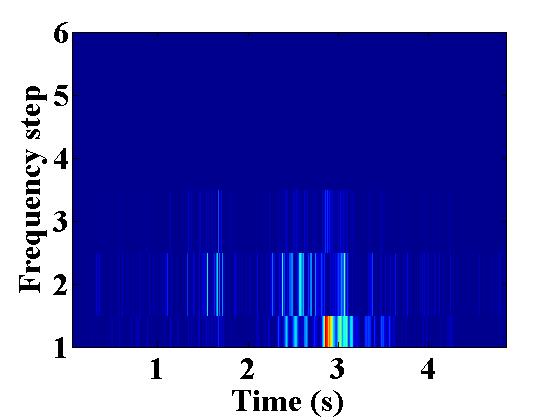}}
\\
\vspace{-6pt}
\centering
\subfigure[PCANet with window size 256]{\includegraphics[width=.45\linewidth,height=2.8cm]{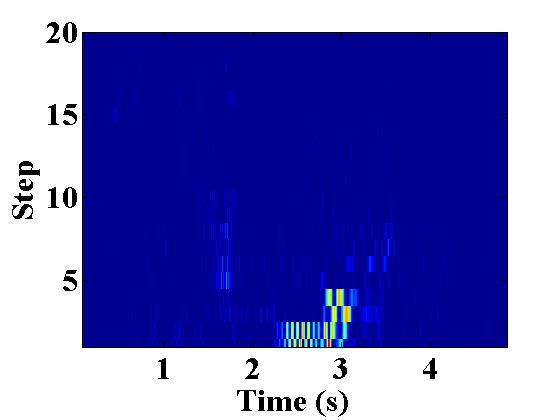}} 
\subfigure[PCANet with window size 20]{\includegraphics[width=.45\linewidth,height=2.8cm]{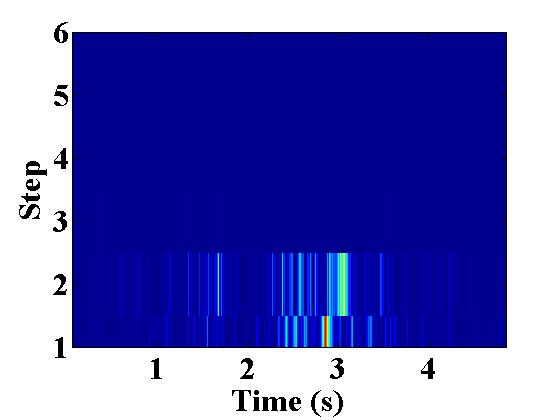}}
\vspace{-6pt}
\caption{Plots of the first layer output}
\label{fig:1st_convolution}
\end{figure}

Plots of short time PCA and DCT (output of the first layer) are shown in Fig.~\ref{fig:1st_convolution} for different window lengths. We use the DCLDE blue whale vocalization data as an example. DCT filterbanks are a natural choice since they are time-frequency representations, whereas time-frequency representation and resolution may be lost when using PCA filterbanks, especially when the PCA eigenfunctions cannot be approximated by the DCT. We can see in the comparison of Fig.~\ref{fig:1st_convolution}(a) and Fig.~\ref{fig:1st_convolution}(c) that PCA fails to represent the time-frequency content of the signal.

The advantage of DCTNet over PCANet can also be found by looking at their computation complexity. PCA is data dependent, and requires $O(M^3)$ operations to find the eigenvectors of the $M\times M$ data covariance matrix, plus an addition computational cost to perform the short time PCA convolution; while for DCTNet, with the help of FFT, we need $O(MN\log_2 M)$ operations to obtain a short time DCT of the signal of length $N$~\cite{ng2015dctnet,wang2012introduction}.

\section{Linear frequency spectrogram}
\label{sec:lfsc}
\begin{figure}[!ht]
\centering
\subfigure[DCTNet 2nd layer output 1]{\includegraphics[width=.45\linewidth,height=2.8cm]{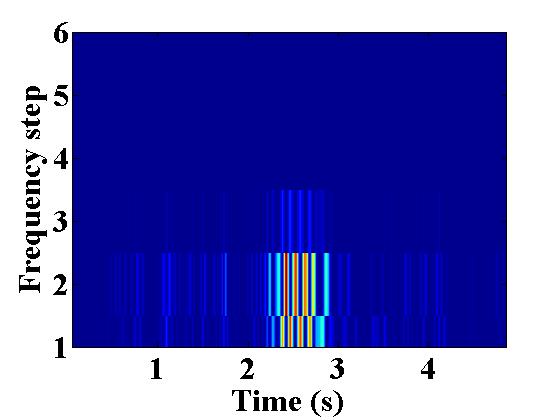}} 
\subfigure[DCTNet 2nd layer output 2]{\includegraphics[width=.45\linewidth,height=2.8cm]{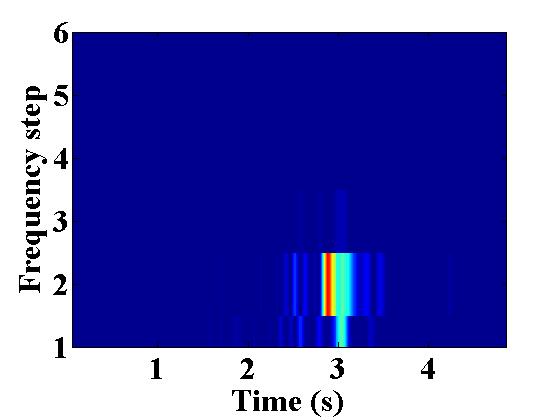}}
\vspace{-6pt}
\caption{DCTNet second layer output with window size 256 at the first layer, and window size 20 at the second layer. (a) shows the signal component at frequency step 14 of Fig.~\ref{fig:1st_convolution}(a), while (b) shows the signal component at frequency step 12 of Fig.~\ref{fig:1st_convolution}(a). }
\label{fig:2nd_convolution}
\end{figure}

After obtaining the first layer, we treat each row of the output as a separate signal, and convolve each with a new PCA or DCT filterbank. We thus end up with multiple new short time PCAs/DCTs, which can capture the dynamic structure of the signal. We choose a smaller window compared with the first layer window size for the filterbanks, so we have a finer scale representation (second layer) inside a coarser scale representation (first layer). Plots of second layer short time DCT are shown in Fig.~\ref{fig:2nd_convolution}. They are generated by taking a short time DCT of window size 20 of the first layer output signal using window size 256, as shown in Fig.~\ref{fig:1st_convolution}(a). Comparing Fig.~\ref{fig:1st_convolution}(b) with Fig.~\ref{fig:2nd_convolution}, we can see that the two-layer DCTNet can illustrate more detailed signal components.

The mel-frequency spectrogram, or MFSC, can be obtained by convolving the signal with a constant-Q filterbank and a low-pass filter, as the analysis in~\cite{anden2011multiscale,anden2014deep}:
\begin{align*}
Mx(t,\lambda)\approx|x*\psi_{\lambda}|^2*|\phi|^2(t),
\end{align*}
where $\psi_{\lambda}$ is a constant-Q band-pass filter, and $\phi$ is a low-pass filter. When replacing the constant-Q filterbank and the low-pass filter with a DCT filterbank and a DCT function, the average of the two-layer DCTNet output can give us a linear frequency spectrogram like features, due to the linearity of scale of DCT filterbank. The average of the two-layer DCTNet features can also improve the deformation stability for acoustic signals.

\begin{figure}[!ht]
\centering
\subfigure[Music log-MFSC feature]{\includegraphics[width=.45\linewidth,height=3cm]{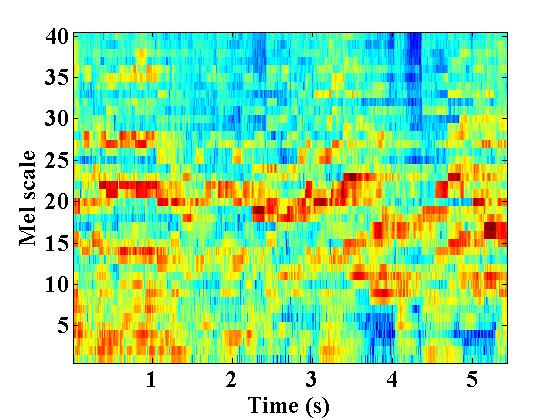}}
\subfigure[$1^{st}$ layer scattering output]{\includegraphics[width=.45\linewidth,height=3cm]{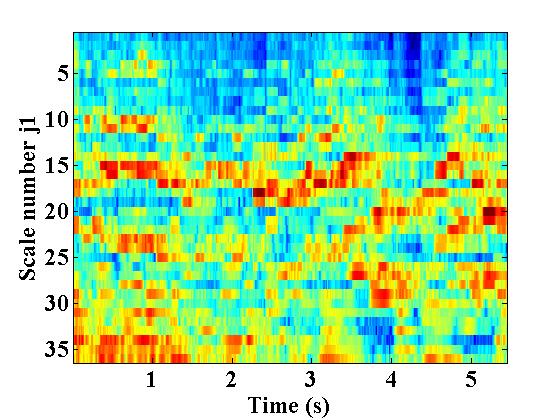}}
\\
\vspace{-6pt}
\centering
\subfigure[Music log-LFSC feature]{\includegraphics[width=.45\linewidth,height=3cm]{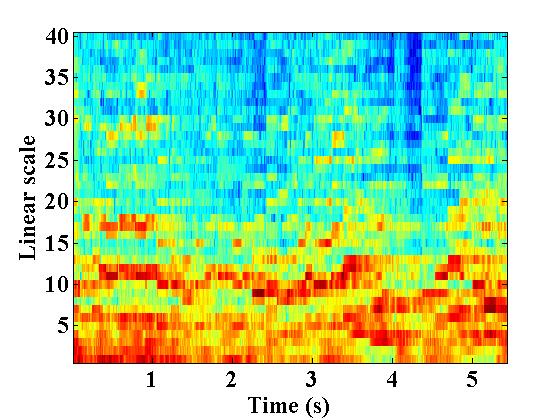}}
\subfigure[Average $2^{nd}$ layer DCTNet]{\includegraphics[width=.45\linewidth,height=3cm]{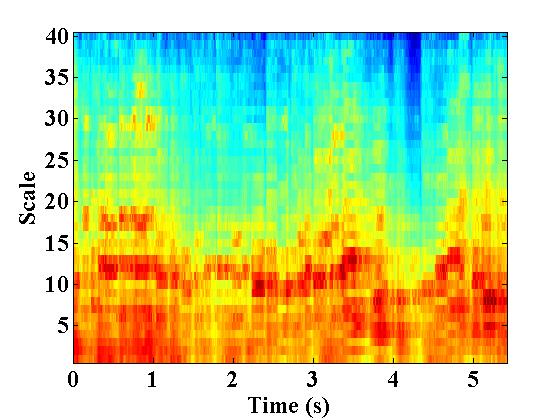}}
\vspace{-6pt}
\caption{Comparison of MFSC, scattering transform coefficients, LFSC and and the average of DCTNet second layer output}
\label{fig:lfsc}
\end{figure}

As Fig~\ref{fig:lfsc} shows, we use a piece of classical music from the GTZAN dataset as an example to illustrate MFSC, LFSC, and their connection with the scattering transform and DCTNet outputs. Applying triangular weight functions for the filterbanks of MFSC and LFSC, we obtain 40 MFSC and LFSC coefficients.  Most of the energy of this music is distributed in frequency range below 4000Hz. The mel-scale is approximately a linear scale for low frequencies~\cite{logan2000mel}, but logarithmic for high frequencies, since the human ear is less sensitive to high frequency signal content~\cite{olson1967music}.




\section{experimental results}
\label{sec:cluster}

\subsection{Dataset}
We use the DCLDE 2015 conference blue whale D call data and fin whale 40Hz call data~\cite{dclde} for experiments. There are 851 blue whale calls and 244 fin whale calls of the same sampling frequency 2000Hz. Each signal lasts for 5 seconds, so the length of each signal is 10000.

\subsection{Spectral Clustering}
We use two-layer DCTNet and PCANet, and use window size 256 for the first layer in order to have a better time-frequency resolution of the data, and choose window size 20 for the second layer to make it comparable to MFSC and LFSC. For MFSC and LFSC, we generate the spectrogram using a Hamming window of length 256. We extract 20 coefficients from each time frame after applying the triangular weighted function filterbank on the signals, and then we concatenate the coefficients along the time axis. We use the MATLAB scattering transform toolbox~\cite{scattering_toolbox} with the frame duration $T=125ms$ (or window size 256) for the experiment, and set $Q_1=12,~Q_2=1$ to better capture the acoustic signal information. In this experiment, the energy of the second layer scattering transform coefficients is very close to zero. Considering the size of dataset and features characteristic, to make a fair experiment, we compared with one layer scattering transform coefficients, MFSC, and LFSC.

\begin{figure}[!ht]
\centering
\subfigure[1-layer DCTNet]{\includegraphics[width=.48\linewidth,height=2.6cm]{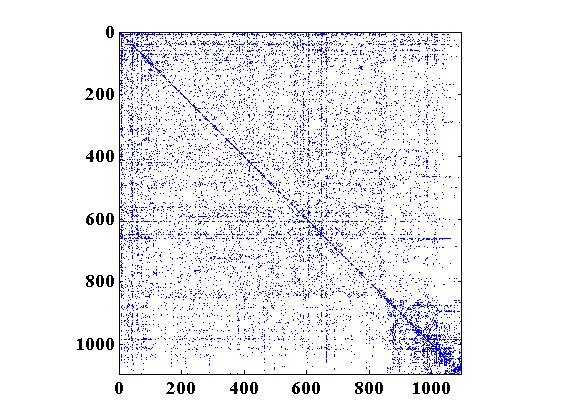}}
\subfigure[2-layer DCTNet]{\includegraphics[width=.48\linewidth,height=2.6cm]{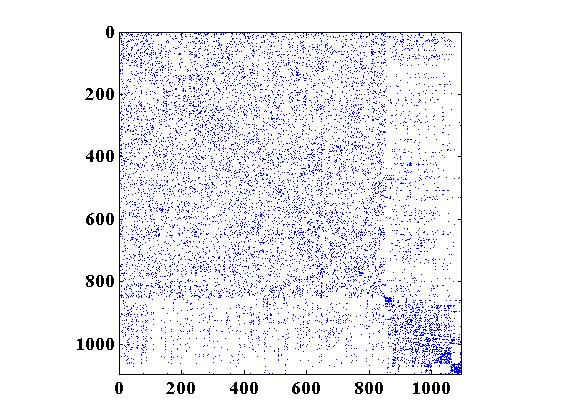}}
\\
\vspace{-6pt}
\centering
\subfigure[1-layer PCANet]{\includegraphics[width=.48\linewidth,height=2.6cm]{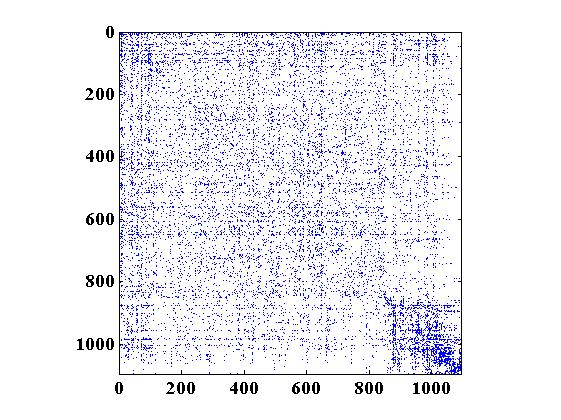}} \subfigure[2-layer PCANet]{\includegraphics[width=.48\linewidth,height=2.6cm]{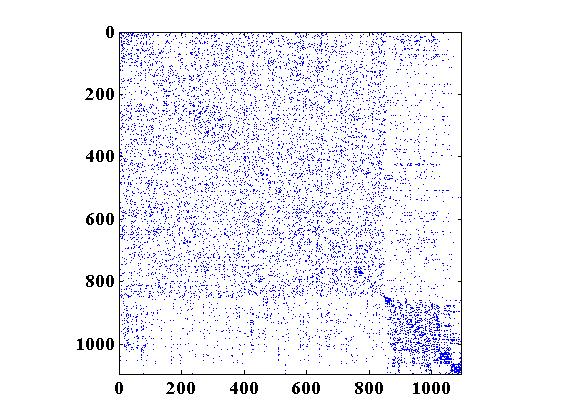}}
\vspace{-6pt}
\caption{Adjacency matrices created by DCTNet and PCANet. The block of rows and columns from 1 to 851 are blue whale data, from 852 to 1095 are fin whale data.}
\label{fig:adj_mat}
\end{figure}

We use the Laplacian Eigenmap for dimensional reduction for the feature set of DCTNet, PCANet, scattering transform, MFSC and LFSC. We use three nearest neighbors to create the kernel of the Laplacian Eigenmap. We examine the adjacency matrices created by the Laplacian Eigenmap, which are generated based on feature similarity.  Since it is a binary classification, we use the Fielder value~\cite{von2007tutorial} for spectral re-ordering. We can see that the two-layer DCTNet and PCANet can better separate the blue whale and fin whale data than one layer, as Fig~\ref{fig:adj_mat} shows.

\begin{figure}[htb]
\includegraphics[height=45mm,width=85mm]{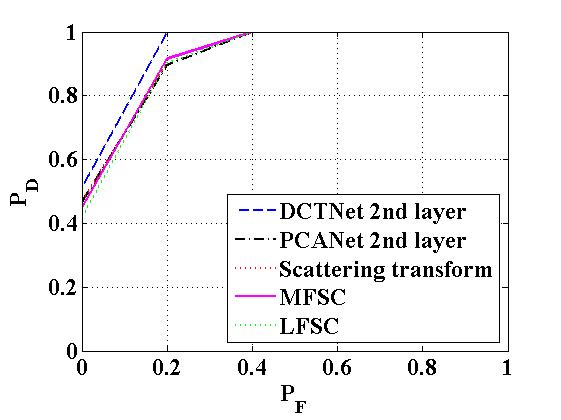}
\vspace{-6pt}
\caption{ROC comparisons}
\label{fig:rocs}
\end{figure}

After obtaining features from DCTNet, PCANet, the scattering transform, MFSC and LFSC, we use the kNN classifier ($k=3$) to evaluate the separation of the data. The ROC plot is shown in Fig.~\ref{fig:rocs}. The values of the AUC (area under the ROC curve) are shown in Table~\ref{table:AUC}.

\begin{table}[h!]
\caption{Area Under the Curves (AUCs) of different feature sets} 
\label{table:AUC}
\centering 
\begin{tabular}{cc}
\hline\hline
Feature set & AUC \\
\hline
DCTNet $2^{nd}$ layer & 0.9513  \\
PCANet $2^{nd}$ layer & 0.9258  \\
DCTNet $1^{st}$ layer & 0.9200   \\
PCANet $1^{st}$ layer & 0.9079   \\
Scattering transform &  0.9275 \\
MFSC & 0.9283\\
LFSC & 0.9225
\\ [1ex]
\hline 
\end{tabular}
\end{table}

We use 5-fold cross-validation to generate the plot, that is 80\% of the blue whale vocalizations and fin whale vocalizations for training, and the rest for testing. We calculate the distance of each testing data point to the training data points, and make a decision based on its three nearest neighbors, and compare it with the ground truth, to obtain the true positive rate and false positive rate. In order to generate the ROC curve, we vary the value of the probability of false alarm ($P_F$) over the range 0 to 1, and obtain a corresponding threshold for the probability of detection $(P_D)$ based upon the obtained true positive and false positive rate. Since the number of blue right whale and fin whale vocalizations available for testing are relatively small, the classification results are promising but preliminary.

\section{Conclusion}
\label{sec:conclude}
In this paper, we apply PCANet and DCTNet for acoustic signal classification. We have shown that each layer of PCANet and DCTNet is essentially a short time PCA/DCT, revealing different time-frequency content of the signal. The computation cost for DCTNet is smaller than for PCANet. DCTNet can generate linear frequency spectral like coefficients, and improve the classification rate of whale vocalizations, which is useful for underwater acoustics, or potentially other applications.



\bibliographystyle{IEEEbib}
\bibliography{refs}

\end{document}